\documentclass[journal]{IEEEtran}
\usepackage{graphicx,amsmath,booktabs,amsfonts,xcolor,cite,makecell,amsthm}
\usepackage[linesnumbered,ruled,vlined]{algorithm2e}
\usepackage[colorlinks,
linkcolor=black,
anchorcolor=black,urlcolor=black,
citecolor=black]{hyperref}
\usepackage{float}
\usepackage[justification=centering]{caption} 

\usepackage{multirow}
\usepackage{multicol}

\ifCLASSINFOpdf
\else
\fi
\begin{document}
%
\title{CBPF: Filtering Poisoned Data Based on Composite Backdoor Attack}
%
%
%

\author{Hanfeng Xia,~\IEEEmembership{}
        Haibo Hong,~\IEEEmembership{}
    and~Ruili Wang~\IEEEmembership{}
\thanks{H. Xia and H. Hong are with School of Computer Science and Technology, Zhejiang Gongshang University, Hangzhou, China. 

R. Wang is with School of Mathematical and Computational Sciences, Massey University, Auckland, New Zealand. 

H. Hong is the corresponding author (email: honghaibo@zjgsu.edu.cn).
}}

%
%

\markboth{IEEE Transactions on Knowledge and Data Engineering,~Vol.~**, No.~**, ***}%
{Shell \MakeLowercase{\textit{et al.}}: Bare Demo of IEEEtran.cls for IEEE Journals}
%



\maketitle

\begin{abstract}
Backdoor attacks involve the injection of a limited quantity of poisoned examples containing triggers into the training dataset. During the inference stage, backdoor attacks can uphold a high level of accuracy for normal examples, yet when presented with trigger-containing instances, the model may erroneously predict them as the targeted class designated by the attacker. This paper explores strategies for mitigating the risks associated with backdoor attacks by examining the filtration of poisoned samples. 
We primarily leverage two key characteristics of backdoor attacks: the ability for multiple backdoors to exist simultaneously within a single model, and the discovery through Composite Backdoor Attack (CBA) that altering two triggers in a sample to new target labels does not compromise the original functionality of the triggers, yet enables the prediction of the data as a new target class when both triggers are present simultaneously. 
Therefore, a novel three-stage poisoning data filtering approach, known as Composite Backdoor Poison Filtering (CBPF), is proposed as an effective solution. Firstly, utilizing the identified distinctions in output between poisoned and clean samples, a subset of data is partitioned to include both poisoned and clean instances. Subsequently, benign triggers are incorporated and labels are adjusted to create new target and benign target classes, thereby prompting the poisoned and clean data to be classified as distinct entities during the inference stage. The experimental results indicate that CBPF is successful in filtering out malicious data produced by six advanced attacks  on CIFAR10 and ImageNet-12. On average, CBPF attains a notable filtering success rate of 99.91\% for the six attacks on CIFAR10. Additionally, the model trained on the uncontaminated samples exhibits sustained high accuracy levels.


\end{abstract}

\begin{IEEEkeywords}
Backdoor Attacks, Backdoor Defenses, Composite Backdoor Poison
Filtering, Deep Neural Network.
\end{IEEEkeywords}

%
\IEEEpeerreviewmaketitle

\section{Introduction}
\label{intro}

\IEEEPARstart{T}{he} rapid development of deep neural networks in various applications, including target detection\cite{1}, image classification\cite{2,3,4}, automatic driving \cite{5}, has yielded promising outcomes and is progressively being implemented in practical settings. As the utilization of neural networks expands, the safeguarding of their models is gaining significance. Recent research \cite{6} has shown that deep learning models are susceptible to backdoor attacks, wherein an attacker strategically associates a chosen trigger with a target label during the model's training phase. This manipulation causes the model to misclassify inputs containing the trigger as the targeted class during inference. The attacker can achieve this by injecting carefully crafted poisoned samples with triggers into the training dataset prior to model training.

To address the risk of backdoor attacks, existing research on backdoor defenses primarily falls into three categories: defense strategies centered on model diagnosis\cite{7,8,9}, sample filtering to detect poisoned samples\cite{10AC,11scan,12strip,13Spectral,14SPECTRE,15asd}, and repression of model poisoning \cite{16abl,17nad,18nab}. Specifically, poisoned sample filtering can be implemented following model diagnosis to identify a compromised model containing a backdoor. For instance, in cases where a model trained on questionable data is identified as containing a backdoor implant, poisoned sample filtering can be employed to segregate the training dataset into tainted and untainted data. While defenses like ABL \cite{16abl} and NAD\cite{17nad} have shown effectiveness in mitigating model poisoning, particularly in repressing backdoors, they often lead to a notable decrease in model accuracy.The latest work NAB\cite{18nab} proposes to use non-adversarial backdoor to suppress the backdoor, although it can get better results but the attacker's backdoor still exists in the model, and the backdoor attack will take effect when the non-adversarial backdoor trigger is missing in the inference phase.

Contrary to defenses centered on the repression of model poisoning, the accurate filtration of all poisoned samples and subsequent training of the model using only clean data can effectively prevent the presence of backdoors and uphold a high level of model accuracy. However, achieving this task is inherently challenging. Early poisoning filtering techniques, exemplified by Strip \cite{12strip}, exhibit improved capability in detecting early backdoor attacks\cite{6} but fall short in completely eliminating poisoned data, often resulting in the loss of clean samples to counter more stealthy backdoor threats \cite{19blend,20bpp,21wanet,22sig,23refool}.
 For SCAn\cite{11scan}, while existing popular backdoor attacks are able to filter most of the poisoned samples with little or no loss of additional clean samples, the remaining portion of the poisoned data is still able to implant a backdoor into the model as the poisoning rate increases.

To solve this problem, we have implemented a novel approach by leveraging the attack strategy utilized in Composite Backdoor Attack 
 (CBA)\cite{24cba} to tackle the challenge of filtering poisoned data. Our method involves utilizing backdoor techniques to effectively segregate the contaminated dataset into two distinct categories: poisoned data and clean data. More specifically, our research introduces a novel training procedure for enhancing backdoor defense, and our method effectively filters out poisoned data without the need for additional clean data, minimizing the loss of clean samples.
%
%
%
%
Specifically, our main contributions are as follows:

\begin{enumerate}
  \item We put forward a novel defense strategy, Composite Backdoor Poison Filtering (CBPF), which effectively separates poisoned data from clean data by leveraging the characteristics of combined backdoors. We adapt CBPF for backdoor defense, resulting in a successful filtration of nearly all poisoned data with minimal impact on clean data.
  \item We test our defense method against six backdoor attacks on two datasets, CIFAR10 and ImageNet-12, and the results demonstrate that CBPF is able to obtain a clean model with high accuracy while removing the backdoor in the original model.
\end{enumerate}


\section{Related Work}
In this section, we will introduce some related works about backdoor attacks and backdoor defences.

\subsection{Backdoor attacks}

Since their inception, backdoor attacks have undergone continuous optimization with respect to three key metrics: increased attack success rate, enhanced attack stealth, and reduced poisoning rate. The introduction of BadNet \cite{6}, which involves the insertion of a black-and-white square into a limited number of training images and manipulation of labels to induce false predictions by the model in the presence of the square, marked a significant milestone in the exploration of backdoor attacks, serving as a catalyst for further research in this area.
This seminal discovery marked the initiation of investigations into backdoor attacks. To enhance stealth, attackers have developed various triggers, including small squares, image pixel blending\cite{19blend}, and subsequently more advanced techniques such as natural reflections \cite{23refool}, invisible noise, backdoor signals \cite{22sig}, image distortions \cite{21wanet}, and pixel quantization\cite{20bpp}. Due to the potential issue of mismatch between image and label information when modifying labels, a clean label attack has been proposed as a method to execute a backdoor attack without altering the label. In addition, recent work \cite{24cba} indicates that backdoor attacks can be realized by combining two different classes of powerful features.

\subsection{Backdoor defenses}

Following the proposal of backdoor attacks,  corresponding defense mechanisms emerged. Through further research on defense strategies, it is observed that poisoned data often exhibit inconsistent behavior compared to clean data within the feature space. This inconsistency allows for the separation of poisoned data from clean data based on differences in feature space representation. Tran et al. \cite{13Spectral} identified that poisoned data exhibits distinct behavior in the covariance spectra of feature representations compared to normal data. This observation was leveraged by Hayase et al. \cite{14SPECTRE} to propose a more efficient filtering method utilizing robust covariance estimation. Additionally, Chen et al.\cite{10AC} employed a clustering approach to segregate poisoned data by creating two clusters for each class. In cases where a class is poisoned, the two clusters exhibit a significant disparity in data volume, facilitating the isolation of the poisoned data.
Tang et al. \cite{11scan} suggested enhancing the clustering method by incorporating the distribution of clean data to enhance clustering effectiveness. Nevertheless, these data separation defense techniques are unable to fully segregate poisoned data from clean data in the presence of more sophisticated backdoor attacks, leading to the presence of backdoors in subsequent training.

In contrast to the distinctions outlined earlier in the feature space, ASD \cite{15asd} represented a dynamic adaptive segmentation dataset defense approach that employs a two-step dynamic filtering process involving loss bootstrapping and meta-learning classification to distinguish between clean and poisoned data. Furthermore, Strip \cite{12strip} advocated for the utilization of predictive entropy in the presence of intentional perturbations to ascertain the presence of poisoned input samples. 

To bolster defenses against backdoor attacks, scholars are integrating backdoor elimination or mitigation strategies into models, expanding upon the concept of data segregation. FP \cite{31fp} found that models exhibit varying neuron activation patterns when predicting poisoned versus clean data. By identifying and selectively removing these activated poisoned neurons, a cleaner model can be obtained. ANP \cite{32ANP} expanded on this concept by demonstrating that exploiting the differences in weight perturbation performance between poisoned and clean neurons can lead to more accurate outcomes. ABL \cite{16abl} also observed that poisoned data is more readily learned.
During the initial phases of training, a limited quantity of tainted data is identified through loss value screening. This subset of tainted data is subsequently used for unlearning on the tainted model post-training. In the context of NAD\cite{17nad}, the acquisition of a pristine model via knowledge distillation with attention mechanisms necessitates the inclusion of additional clean samples, thereby introducing additional limitations when implementing such solutions in practical scenarios.

In addition to utilizing unlearning or distillation with poisoned data, NAB\cite{18nab} found that it's possible to add its own trigger on poisoned data and attempt to revert the labels back to the correct ones as much as possible for training the model. In this way, during the inference stage, it can use its own trigger to repress the backdoor in the model.

This paper diverges from the approach taken by NAB, which requires the insertion of non-adversarial triggers into poisoned data and the reassignment of labels to their original values in order to mitigate backdoor effects. Furthermore, although the filtering method used in NAB can filter out most of the poisoned data, there will still be many instances of poisoned data remaining in the clean data, especially at higher poisoning rates, which can still achieve backdoor effects. NAB primarily addresses poisoning repression in model, thereby allowing the backdoor to persist in the compromised model, thus perpetuating security vulnerabilities. Prior to training the model, our proposed method capitalizes on the properties of composite backdoors to differentiate poisoned data from clean data. This approach enables the development of a completely backdoor-free model, utilizing only clean data during the training process.

\section{Poisoning Data Filtering for Combination Triggers}
\label{section2}
\subsection{Threat Model}
It is posited that the malicious attacker responsible for the backdoor attack has embedded poisoned data with triggers into the training dataset. Additionally, it is assumed that the defender lacks any prior knowledge regarding the target class, attack methodology, or trigger style employed by the attacker. Despite this, the defender retains control over the entire training process with the objective of developing a pristine model on the compromised training set that can accurately predict clean samples. The goal of the defender is to train a clean model on the suspect training set that maintains high prediction accuracy on clean samples. This scenario is more in line with actual training scenarios.

\subsection{Problem Formulation}
On a standard dataset used for image classification, an attacker selects a small number $n_p$ of clean samples from the original dataset $D$ to create poisoned samples by adding a trigger $\overline{x}=x\oplus\delta$, where $\delta$ is the trigger and the label $y$ corresponding to $x$ is modified to the attacker's target label $y_{\text{target}}$. The remaining dataset is used as a clean dataset $D_{c}=\left\{\left(x_{i}, y_{i}\right)\right\}_{i=1}^{n_{c}}$. Then, the set of poisoned data $D_{p}=\left\{\left(\bar{x_i}, y_{\text{target }}\right)\right\}_{i=1}^{n_{p}}$ and the clean dataset $D_{c}$ are merged to form the training set $\bar{D}=D_{p} \cup D_{c} $. The standard training process to train a model is formulated as follows:
\begin{equation}
\mathcal{L}=E_{(x, y) \sim \bar{D}}\left[\ell\left(f_{\theta}(x), y\right)\right],
\end{equation}
where $\ell$ denotes  cross-loss entropy function, $f_\theta$ denotes the classification function of the model. Since we can represent the whole dataset as two parts $D_p$ and $D_c$,  this loss function is further defined as:
\begin{equation}
\mathcal{L}=E_{(x, y) \sim D_{c}}\left[\ell\left(f_{\theta}(x), y\right)\right]+E_{\left(\bar{x}, y_{\text {target }}\right) \sim D_{p}}\left[\ell\left(f_{\theta}(\bar{x}), y_{\text {target }}\right)\right].
\end{equation}

The efficacy of the backdoor attack hinges primarily on the latter portion of formula (2), which involves the establishment of a robust correlation between the trigger and the target label through the prediction of ``shortcuts" between them. Consequently, upon the presence of the trigger in the input data, $f_{\theta}(\bar{x})$ will reliably predict it as $y_{\text{target}}$.

In the integrated backdoor defense strategy, the normal model is trained by selectively filtering out poisoned data and exclusively utilizing clean data. This process ensures that the model exclusively learns the correlation between clean images and their corresponding correct labels in the initial portion of Equation 2. Consequently, the filtered poisoned data prevents the model from establishing any association between the trigger and the target label.

\subsection{Differences in output of poisoned samples}

After introducing a trigger to manipulate data in a model with a backdoor, the model consistently classifies that data as the target class. Building on this observation, we hypothesize that models with backdoors designed to classify poisoned data exhibit significantly higher output values for the target class compared to other classes, even when the correct class aligns with the image information. To quantify this discrepancy more effectively, we propose a straightforward metric-calculating the difference between the top two maximum output values of the model, denoted as $\Delta$$\text{Top}2_{\text{diff}}$:
\begin{equation}
    \Delta \text{Top}2_{\text {diff }}=\left|\max \left(f_{\theta}(x)\right)-2^{\text {nd }} \max \left(f_{\theta}(x)\right)\right|\label{eq:top2diff},
\end{equation}
where $f_{\theta}(x)$ is the output value of the function $f_\theta$ for the input $x$, which represents the predicted value of $x$ for all categories, and $\max (\cdot)$, $2^{\text{nd}}\max (\cdot)$ means to take the maximum and the second largest value, respectively. Here, the larger the value of $\Delta$$\text{Top}2_{\text{diff}}$, the more likely it is to be a poisoned sample.



\subsection{Method}

This section provides a formal introduction to the workflow of CBPF, which aims to differentiate between poisoned and clean data by manipulating them into separate target classes through the use of benign triggers and label modifications.

\begin{figure*}[htbp]
	\centering
	\includegraphics[width=\textwidth]{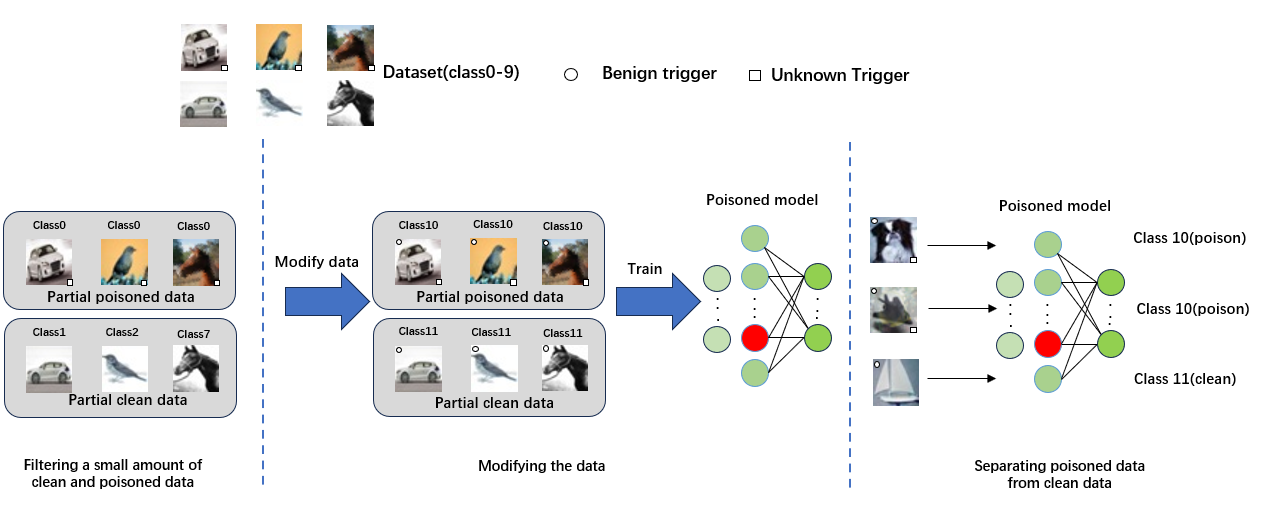}   \captionsetup{singlelinecheck=off,justification=justified}
	\caption{Overview of the proposed framework. (A) Filtering a small amount of clean and poisoned data: The defender initially obtains a limited number of clean and poisoned samples through filtration. (B) Modifying the data: The defender augments the clean data with benign triggers and reassigns the labels to benign in order to enhance the benign target class. Additionally, the defender applies the same benign trigger to the poisoned data and changes the label to the new target class to execute the Composite Backdoor Attack (CBA). (C) Separating poisoned data from clean data: During the testing phase, the defender applies benign triggers to each input. Inputs predicted as a new target class are filtered out as poisoned data, while those predicted as benign target classes are retained as clean data.}
	\label{fig:2}
\end{figure*}

Specifically, in the context of composite backdoor attacks, the presence of two distinct strong features from disparate classes within a sample can be leveraged to retrain a model to classify the sample into a new target label, thereby simultaneously showcasing both strong features. It is observed that the trigger attributes of the backdoor attack align closely with the robust characteristics of the attacker's designated target class. Afterwards, we introduce our own triggers into the clean data as potent features of the benign target category and subsequently adjust them to align with the benign target category. The triggers incorporated within the poisoned data exhibit significant characteristics of their respective target categories. Consequently, a composite backdoor attack may be executed by introducing benign triggers into the isolated poisoned data. This method of composite backdoor attack does not disrupt the correlation between these significant features and their original categories. As a result, all remaining clean images with benign triggers appended will be classified as benign target categories. This approach effectively segregates the poisoned data from the clean data. As shown in Fig.\ref{fig:2} and Algorithm \ref{algorithm1}, CBPF can be divided into three steps.

\textbf{Filtering a small amount of clean and poisoned data.}
The training set $\bar{D}$ is duplicated to create the training set $D'$. Subsequently, the model $f_{\text{filter}}$ is trained on the $D'$ training set to facilitate the segregation of the training sets. The $\Delta$$\text{Top}2_{\text{diff}}$ metric is computed for each datum in the dataset by applying Equation \ref{eq:top2diff}.
Following the sorting of data from largest to smallest based on a specific value, as demonstrated in previous research by ABL\cite{16abl}, it has been observed that the model tends to prioritize the acquisition of associations between triggers and target labels during the initial epochs when the data is perturbed. Consequently, we proceed to compute the $\Delta$$\text{Top}2_{\text{diff}}$ metric for each input following an early training phase, typically after 5 epochs. The small subset of data exhibiting the highest $\Delta$$\text{Top}2_{\text{diff}}$ value, such as 3\% in CIFAR10 \cite{25cifar10}, is classified as poisoned data due to the expectation that following initial training, the model will assign a significantly higher probability to the target label compared to the second highest prediction. Conversely, a subset of data with the lowest $\Delta$$\text{Top}2_{\text{diff}}$ values, for example 1\% in CIFAR10, is considered clean data.

\textbf{Modifying the data.} Benign triggers are incorporated into the poisoned data within the  training set $D'$, with corresponding modifications made to the labels to reflect the new target labels. To prevent interference with other normal classes, an additional target class is introduced (e.g., in CIFAR10, labeled as class 0-9, label 10 is designated as the new target class). Similarly, benign triggers are added to the clean data within $D'$, resulting in a new class and modification of the label to reflect a benign target class (e.g., in CIFAR10, label 11 is designated as the benign target class). Subsequently, all training sets within $D'$ are utilized to train a poisoning model.

\textbf{Separating poisoned data from clean data.} All data within the  training set $D'$ are augmented with benign triggers, and subsequently predicted by the poisoned model derived from the second step of training. Data classified as belonging to a new target class are allocated to the poisoned data pool, while those classified as benign target class are allocated to the clean data pool. The identification of corresponding data within $\bar{D}$ that has not been augmented with benign triggers is then conducted within the poisoned data pool and clean data pool. A new clean model is subsequently trained using the data from the clean data pool.

\SetKwInput{KwInput}{Input}                
\SetKwInput{KwOutput}{Output}              
\SetKwInput{KWHyperparameters}{Hyper-parameters} 
\begin{algorithm}[htbp]
	\DontPrintSemicolon
	\KwInput{\Indm\\${D'}$: backdoor-poisoned dataset backup;$D_{p}^{\prime}$: filtered portion of poisoned data; $D_{c}^{\prime}$: filtered portion of clean data; $\ell(\cdot)$: cross-entropy loss;$f_\theta$: Initialized model function; }
	\KwOutput{\Indm\\$f_{\text {filter}}$: a model for separating poisoning and clean data; $N$:number of filter;}
        \KWHyperparameters {\Indm\\$T_{\text {early}}$: training epoch for the filter model; $y_{\text {new\ target}}$:  new target class; $y_{\text {benign}}$: benign target class; ${\ a}_{p\ }$: poisoning isolation rate ; ${\ a}_{c\ }$: clean data isolation rate; }
        \For{$i$ in $[1, \ldots, T_{\text {early}}]$}
{
    $f_{\text {filter}} \leftarrow \operatorname{argmin} E_{(x, y) \sim \bar{D}}\left[\ell\left(f_{\theta}(x), y\right)\right]$ \\
}
end for

\#Calculate difference

        \For {$x$ in $\overline{D}$}
{
		$\Delta=\frac{\sum_{k}^{N}\left|\max \left(f_{ \text{ filter}}(x)\right)-2^{\text{nd}} \max \left(f_{\text{filter}}(x)\right)\right|}{N}$\\
}
end for

sorted($\overline{D}$, $\Delta$$\text{Top}2_{\text{diff}}$)

$D_{p}^{\prime} \leftarrow \text { The } a_{p} \text { largest } \Delta \text { samples in } \bar{D}$

$D_{c}^{\prime} \leftarrow \text { The } a_{c} \text { smallest } \Delta \text { samples in } \bar{D}$

\For{$(x,y)$ in $D_{p}^{\prime}$}
{
    $x=x \oplus \delta_{\text {benign}}$

    $y=y_{\text {new target}}$
}
end for

\For{$(x,y)$ in $D_{c}^{\prime}$}
{
    $x=x \oplus \delta_{\text {benign }}$

    $y=y_{\text {benign target }}$
}
end for

	\caption{Data manipulation steps for CBPF}
	\label{algorithm1}
\end{algorithm}

\section{EXPERIMENTS}	
\label{section3}

\subsection{Attack Configurations}
\begin{figure}[htbp]
    \centering
    \includegraphics[width=0.5\textwidth]{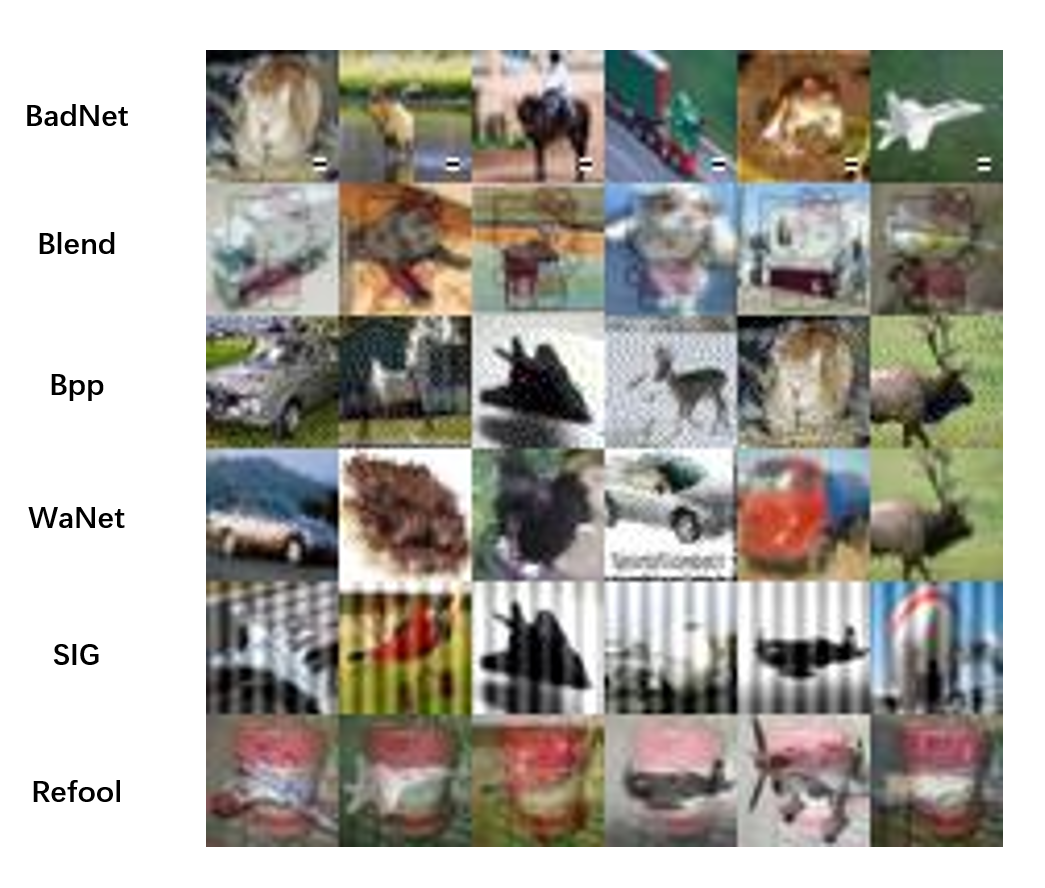}
    \caption{Examples of backdoor poisoning attacks in our experiments.}
    \label{fig:222}
\end{figure}

We consider six backdoor attacks including four dirty label attacks: BadNet\cite{6}, image blending attack (Blend)\cite{19blend}, image quantization attack (BPP)\cite{20bpp}, distortion based attack (WaNet)\cite{21wanet}, two clean label attacks: sinusoidal signal attack (SIG)\cite{22sig}, natural reflection attack (Refool)\cite{23refool}. Table \ref{table2} summarizes the specific details of all the backdoor triggers. Fig \ref{fig:222} llustrates the specific effects of different attacks. We implement these attacks based on the settings suggested by the papers in these attacks and their open source code. We use the ResNet18\cite{27resnet} model for evaluation on CIFAR10 \cite{25cifar10} and ImageNet-12\cite{26imagenet}. Since the WaNet paper does not provide an attack on ImageNet-12, we evaluate the attack only on CIFAR10. Here, we set the poisoning rate of both the dirty label attack to 10\%, the clean label attack to 7\%, and the attacker's target label to label 0. In the subsequent experiments, we mainly conduct them on CIFAR10. The datasets  used in our experiments are summarized in Table \ref{tab:dataset_comparison}.

\begin{table}[htbp]
\caption{DETAILS OF DATASETS AND CLASSIFIERS USED IN THE EXPERIMENT.}
\label{tab:dataset_comparison}
\begin{tabular}{|c|c|c|c|c|}
\hline
Dataset     & \multicolumn{1}{l|}{Classes} & \multicolumn{1}{l|}{Input Size} & \multicolumn{1}{l|}{Train Number} & \multicolumn{1}{l|}{Tested Data} \\ \hline
CIFAR10    & 10                           & 32×32×3                         & 50000                             & 10000                            \\ \hline
ImageNet-12 & 12                           & 224×224×3                       & 5760                              & 1440                             \\ \hline
\end{tabular}
\end{table}

\begin{table*}[htbp]
	\begin{center}
		\caption{Attack settings for six backdoor attacks.}
		\setlength{\tabcolsep}{8mm}{
			\begin{tabular}{ccccccc}
				\toprule  
				Attack& Trigger Type& Trigger Pattern& Target& Poisoning Rate\\
				\midrule  
				BadNets& Fixed& Square& 0& 10\%\\
				Blend& Fixed& Hello Kitty& 0& 10\%\\
				Bpp& Varied& Noise Disturbance& 0& 10\%\\
				WaNet& Varied& Warping& 0& 10\%\\
				SIG& Fixed& Sinusoidal Signal& 0& 7\%\\
				Refool& Fixed&Reflection& 0& 7\%\\
				
				\bottomrule 
			\end{tabular}
		}
		\label{table2}
	\end{center}
\end{table*}

\subsection{Defense and Training Details}

We conduct a comparative analysis of our defense method against four contemporary defense methods. Specifically, we evaluat two techniques for training clean models on poisoned datasets: Neural Attention Distillation (NAD)\cite{17nad}  and Anti-Backdoor Learning (ABL)\cite{16abl}, as well as two data filtering methods known as Strip \cite{12strip} and SCAn\cite{11scan}. 

In our evaluation of NAD, we utilize the publicly available code from the original paper as a foundation. Given the disparities between the model architecture in our experiments and the model employed in NAD, we make meticulous adjustments to the $\beta$-values to optimize performance in our model. We fine-tune the poisoning model for 10 epochs using 5\% clean data according to the settings in the paper. In CIFAR10, we train 70 epochs in the teacher model versus the instructor student model with $\beta_1$=0, $\beta_2$=0, $\beta_3$=1000, $\beta_4$=1000.

In accordance with the methodology outlined in the paper, the ABL model undergoes training for 20 epochs during the initial warm-up phase to filter out 1\% of the poisoning data with a parameter value of $\gamma$ = 0.5. Subsequently, following 60 epochs of training on the complete dataset, unlearning is conducted during the final 20 epochs using 1\% of the poisoning data and a learning rate of 0.0001.

The optimization procedures for Strip and SCAn models adhere to the code implementation specified in the original publication.

The defensive strategy proposed is elucidated in Algorithm 1. On the CIFAR10 dataset, we use the full data to train the filtering model on the model used for screening, $T_{\text{early}}$ =5,${\ a}_{p\ }$=3\% , ${\ a}_{c\ }$=1\%. On the ImageNet-12 dataset, $T_{\text{early}}$ =30,${\ a}_{p\ }$=3\% , ${\ a}_{c\ }$=10\%. On both datasets we use the SGD optimizer with learning rate =0.1 and adjust the learning rate to 0.1 times the previous one after 50, 75 epochs of training.

\begin{table*}[htbp]
\centering
\caption{The attack success rate (ASR) and accuracy of clean accuracy rate (CAR) of five backdoor defense methods against six backdoor attacks.}
\label{compare other defense}
\begin{tabular}{l|cclclclclclclc}
\hline
\multicolumn{1}{c|}{(\%)} & Attack               &     &  & No Defense     &  & Strip &  & SCAn           &  & ABL           &  & NAD           &  & CBPF(Ours)          \\ \hline
                          & \multirow{2}{*}{BadNet}               & CAR &  & 93.34          &  & 92.60 &  & \textbf{93.46} &  & 90.67         &  & 84.36         &  & 93.38          \\
                          &                      & ASR &  & 100            &  & 1.13  &  & 0.80           &  & \textbf{0.09} &  & 1.61          &  & 0.82           \\ \cline{2-15} 
                          & \multirow{2}{*}{Blend}                & CAR &  & 93.94          &  & 92.70 &  & \textbf{93.48} &  & 85.26         &  & 84.27         &  & 93.38          \\
                          &                      & ASR &  & 99.94          &  & 99.82 &  & 3.99           &  & 1.48          &  & \textbf{0.53} &  & 2.49           \\ \cline{2-15} 
                          & \multirow{2}{*}{Bpp}                  & CAR &  & 93.57          &  & 92.99 &  & \textbf{93.38} &  & 86.64         &  & 85.57         &  & 93.10          \\
                          &                      & ASR &  & 100            &  & 0.89  &  & 12.18          &  & \textbf{0.39} &  & 0.87          &  & 3.82           \\ \cline{2-15} 
CIFAR10                   & \multirow{2}{*}{WaNet}                & CAR &  & 93.33          &  & 92.67 &  & \textbf{93.37} &  & 86.76         &  & 88.30         &  & 93.17          \\
                          &                      & ASR &  & 100            &  & 99.23 &  & 99.91          &  & \textbf{0.08} &  & 1.16          &  & 2.33           \\ \cline{2-15} 
                          & \multirow{2}{*}{SIG}                  & CAR &  & 93.33          &  & 93.46 &  & 92.83          &  & 81.48         &  & 82.96         &  & \textbf{93.63} \\
                          &                      & ASR &  & 98.07          &  & 0.23  &  & 0.11           &  & 0.54          &  & 0.12          &  & \textbf{0.08}  \\ \cline{2-15} 
                          & \multirow{2}{*}{Refool}               & CAR &  & 94.05          &  & 92.51 &  & 92.60          &  & 84.82         &  & 86.11         &  & \textbf{93.66} \\
                          &                      & ASR &  & 97.18          &  & 93.56 &  & 13.89          &  & \textbf{0.01} &  & 2.07          &  & 0.21           \\ \hline
\multicolumn{1}{c|}{}     & \multirow{2}{*}{BadNet}               & CAR &  & 84.24          &  & 83.03 &  & \textbf{84.55} &  & 82.05         &  & 76.06         &  & 83.64          \\
                          &                      & ASR &  & 94.45          &  & 1.59  &  & 1.59           &  & \textbf{0.45} &  & 1.97          &  & 1.36           \\ \cline{2-15} 
                          & \multirow{2}{*}{Blend}                & CAR &  & 85.42 &  & 83.18 &  & 83.63          &  & 75.76         &  & 73.86         &  & \textbf{85.38}         \\
                          &                      & ASR &  & 99.55          &  & 93.18 &  & 68.64          &  & 7.50          &  & 33.18         &  & \textbf{1.44}  \\ \cline{2-15} 
ImageNet-12               & \multirow{2}{*}{Bpp}                  & CAR &  & 85.00          &  & 84.17 &  & 83.03          &  & 69.70         &  & 69.70         &  & \textbf{85.76} \\
                          &                      & ASR &  & 99.70          &  & 2.05  &  & 6.82           &  & 4.70          &  & 4.24          &  & \textbf{0.30}           \\ \cline{2-15} 
                          & \multirow{2}{*}{SIG}                  & CAR &  & 87.20 &  & 82.05 &  & 85.45          &  & 61.74         &  & 72.12         &  & \textbf{87.05}          \\
                          &                      & ASR &  & 89.09          &  & 83.86 &  & 12.80          &  & \textbf{0}    &  & 5.68          &  & 0.08           \\ \cline{2-15} 
                          & \multirow{2}{*}{Refool}               & CAR &  & 86.06          &  & 83.03 &  & 85.91          &  & 76.06         &  & 72.35         &  & \textbf{86.82} \\
                          & \multicolumn{1}{l}{} & ASR &  & 95.61          &  & 95.76 &  & 77.80          &  & 1.97          &  & 19.70         &  & \textbf{0.23}  \\ \hline
\end{tabular}
\end{table*}

\subsection{Metrics}
Two conventional metrics are utilized in backdoor defense: Attack Success Rate (ASR), which measures the precision of being identified as the target class when evaluating the model on poisoned test data, and Clean Accuracy Rate (CAR), which assesses the model's prediction accuracy on clean data. Furthermore, two parameters have been defined: the True Positive Rate (TPR), which quantifies the proportion of poisoned data correctly identified within the poisoned dataset, and the False Positive Rate (FPR), which indicates the percentage of clean samples inaccurately classified as poisoned among the total clean samples.

\begin{table}[htbp]
\caption{The filtering effectiveness of three poisoning filtering methods against six
different backdoor attacks.
}
\label{Compare the filtering}
\begin{tabular}{l|ccclclc}
\hline
\multicolumn{1}{c|}{\%}       & Attack                  &     & Strip          &           & SCAn                      &           & CBPF(Ours)     \\ \hline
                              & \multirow{2}{*}{BadNet} & TPR & \textbf{100}   &           & \textbf{100}              &           & \textbf{100}   \\
                              &                         & FPR & 9.69           &           & \textbf{0}                &           & 0.04           \\ \cline{2-8} 
                              & \multirow{2}{*}{Blend}  & TPR & 83.12          &           & 99.69                     &           & \textbf{99.92} \\
                              &                         & FPR & 9.61           &           & \textbf{0}                &           & \textbf{0}     \\ \cline{2-8} 
                              & \multirow{2}{*}{Bpp}    & TPR & \textbf{100}   &           & 99.52                     &           & 99.92          \\
                              &                         & FPR & 9.79           &           & \textbf{0}                &           & \textbf{0}     \\ \cline{2-8} 
CIFAR10                       & \multirow{2}{*}{WaNet}  & TPR & 85.74          &           & \multicolumn{1}{l}{61.72} &           & \textbf{99.66} \\
                              &                         & FPR & 9.61           &           & \textbf{0}                &           & 0.11           \\ \cline{2-8} 
                              & \multirow{2}{*}{SIG}    & TPR & 99.97          &           & 99.91                     &           & \textbf{100}   \\
                              &                         & FPR & 9.39           &           & 5.44                      &           & \textbf{0}     \\ \cline{2-8} 
                              & \multirow{2}{*}{Refool} & TPR & 0.71           &           & 98.05                     &           & \textbf{99.89} \\
                              &                         & FPR & 9.61           &           & \textbf{0}                &           & 0.03           \\ \hline
\multirow{10}{*}{ImageNet-12} & \multirow{2}{*}{BadNet} & TPR & \textbf{99.30} &           & 93.05                     &           & 98.96          \\
                              &                         & FPR & 11.25          &           & \textbf{0.03}             & \textbf{} & 2.25           \\ \cline{2-8} 
                              & \multirow{2}{*}{Blend}  & TPR & 72.39          &           & 92.88                     &           & \textbf{99.65} \\
                              &                         & FPR & 7.77           &           & \textbf{0.15}             & \textbf{} & 6.01           \\ \cline{2-8} 
                              & \multirow{2}{*}{Bpp}    & TPR & 99.47          & \textbf{} & 98.95                     &           & \textbf{100}   \\
                              &                         & FPR & 9.02           &           & 4.08                      & \textbf{} & \textbf{3.74}  \\ \cline{2-8} 
                              & \multirow{2}{*}{SIG}    & TPR & 89.33          &           & 97.27                     &           & \textbf{100}   \\
                              &                         & FPR & 13.42          &           & \textbf{0}                &           & 1.92           \\ \cline{2-8} 
                              & \multirow{2}{*}{Refool} & TPR & 0              &           & 97.02                     &           & \textbf{99.89} \\
                              &                         & FPR & 12.91          &           & \textbf{0.09}             & \textbf{} & 6.60           \\ \hline
\end{tabular}
\end{table}

\subsection{Comparison to Existing Defenses}

The efficacy of our methodology in mitigating six established advanced backdoors is demonstrated in Table \ref{compare other defense}, juxtaposed with two traditional backdoor removal defense strategies and two data filtering methods for poison data. As defenders, our objective is to assess the experimental impact of the model's ability to maintain a comparable accuracy rate of backdoor defense, while observing a significant decrease in the attack success rate. Conversely, in the context of poisoned data filtering, our dual objectives are to ensure a high TPR and minimize the FPR.

\textbf{Comparison with Strip.} The results presented in Table I demonstrate that CBPF outperforms Strip in defending against various attacks. Specifically, CBPF successfully mitigates backdoor attacks on the CIFAR10 dataset while maintaining a high model accuracy, with an average accuracy of 93.39\% and an average ASR of 1.62\%. In contrast, Strip exhibits lower performance in defending against certain attacks, leading to a higher average ASR of 49.14\% despite a comparable average accuracy of 92.82\% to CBPF. Notably, Strip's vulnerability to attacks such as WaNet and Refool contributes significantly to its higher ASR compared to CBPF.
The failure of Strip in effectively defending against backdoor attacks can be attributed to its reliance on backdoor triggers for defense. In comparison to WaNet and Refool's backdoor attacks, Strip's backdoor mechanism is weaker, resulting in a higher number of poisoned samples evading detection during data separation, as outlined in Table \ref{Compare the filtering}. Despite CBPF demonstrating greater efficacy in preserving backdoor integrity on the WaNet dataset within ImageNet-12, it still surpasses Strip in maintaining model accuracy (85.73\% vs. 83.09\%) and successfully eliminating backdoor tasks (0.68\% vs. 55.29\%).

\textbf{Comparison with SCAn.} As demonstrated in Table \ref{compare other defense}, CBPF exhibits superior performance compared to SCAn in the context of backdoor elimination, particularly on the ImageNet-12 dataset. Specifically, on the CIFAR10 dataset, CBPF demonstrates a 20.19\% lower ASR than SCAn (1.62\% vs. 21.81\%), while maintaining a comparable CAR to SCAn (93.39\% vs. 93.18\%). Moreover, on the ImageNet-12 dataset, CBPF's superiority in both CAR (85.73\% vs. 84.51\%) and ASR (1.62\% vs. 33.53\%) is even more evident.

\textbf{Comparison with ABL.} According to the data presented in Table \ref{compare other defense}, both CBPF and ABL demonstrate superior performance in the backdoor elimination task. Specifically, ABL achieved a notable reduction in the average ASR on the CIFAR10 dataset, decreasing from 99.19\% without defense to 0.43\%, although this improvement was accompanied by a more substantial decrease in CAR from 93.59\% to 85.93\%. In contrast, CBPF, while maintaining a comparable ASR to ABL (1.62\% vs. 0.43\%), exhibited a higher CAR of 93.39\% compared to ABL's 85.93\%. Furthermore, CBPF consistently outperformed ABL on ImageNet-12 in maintaining model integrity, with a CAR of 85.73\% compared to ABL's 74.06\%. We believe that the decline in clean accuracy of the ABL model can be attributed to the integration of attack triggers such as Blend and Refool with the original image, leading to the mixing of backdoor and clean functions within the model. During the process of unlearning this poisoned functionality, a portion of the clean functionality may inadvertently be forgotten.

\textbf{Comparison with NAD.} As illustrated in Table \ref{compare other defense}, the analysis reveals that both CBPF and NAD exhibit superior performance in the context of backdoor removal. Specifically, on CIFAR10, NAD demonstrates a reduction in average clean accuracy by 8.33\% (from 93.59\% to 85.26\%) while effectively eliminating the backdoor (from 99.19\% to 1.06\%) compared to the pre-defense state. In contrast, CBPF achieves a higher average clean accuracy of 8.13\% (93.3\% versus 85.26\%) compared to NAD. Furthermore, the superiority of CBPF over NAD is evident in both average clean accuracy (85.73\% versus 72.81\%) and backdoor removal efficacy (1.62\% versus 12.95\%) on ImageNet-12.

Additionally, we conduct a comparison of the filtering efficacy of CBPF with two poisoning filtering techniques, as illustrated in Table \ref{Compare the filtering}. Our analysis reveals that on CIFAR10, both Strip and Scan exhibit a TPR of less than 80\% for at least one type of attack, indicating a lack of effectiveness in filtering. In contrast, CBPF demonstrates a commendable average TPR exceeding 99.89\% across all attacks. Furthermore, the FPR of CBPF defense stands at a mere 0.03\%, comparable to that of SCAn. On the ImageNet-12 dataset, CBPF demonstrates a consistently high TPR, surpassing Strip by an average of 27.69\% (99.7\% vs. 72.01\%) and SCAn by 3.87\% (99.7\% vs. 95.83\%). This observation highlights the efficacy of employing a combined backdoor approach in effectively filtering out poisoned data, as evidenced by CBPF successfully filtering the majority of tainted data in both datasets.

\begin{figure*}[htbp]
    \centering
    \begin{minipage}[b]{0.5\textwidth}
        \centering
        \includegraphics[width=\linewidth]{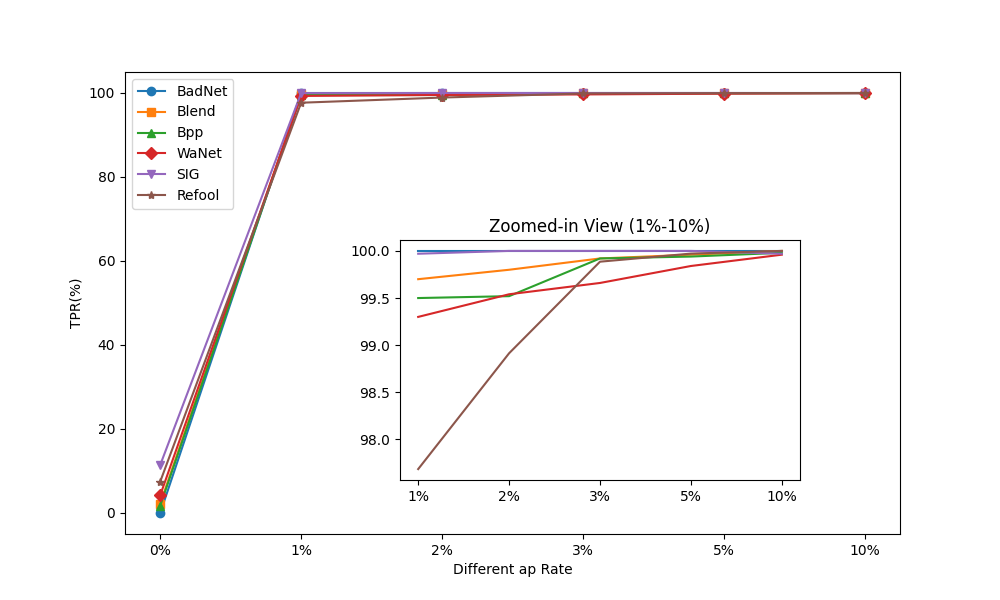}
        \caption{Performance of TPR under fixed $a_c$ in different $a_p$.}
        \label{fig:1}
    \end{minipage}%
    \begin{minipage}[b]{0.5\textwidth}
        \centering
        \includegraphics[width=\linewidth]{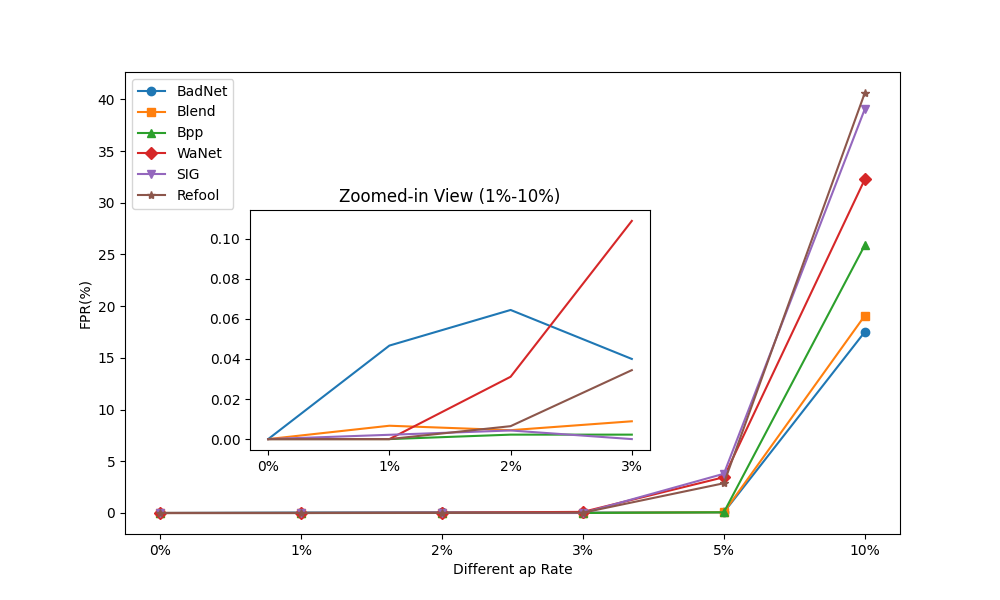}
        \caption{Performance of FPR under fixed $a_c$ in different $a_p$.}
        \label{fig:22}
    \end{minipage}
\end{figure*}

\begin{figure*}[htbp]
    \centering
    \begin{minipage}[b]{0.5\textwidth}
        \centering
        \includegraphics[width=\linewidth]{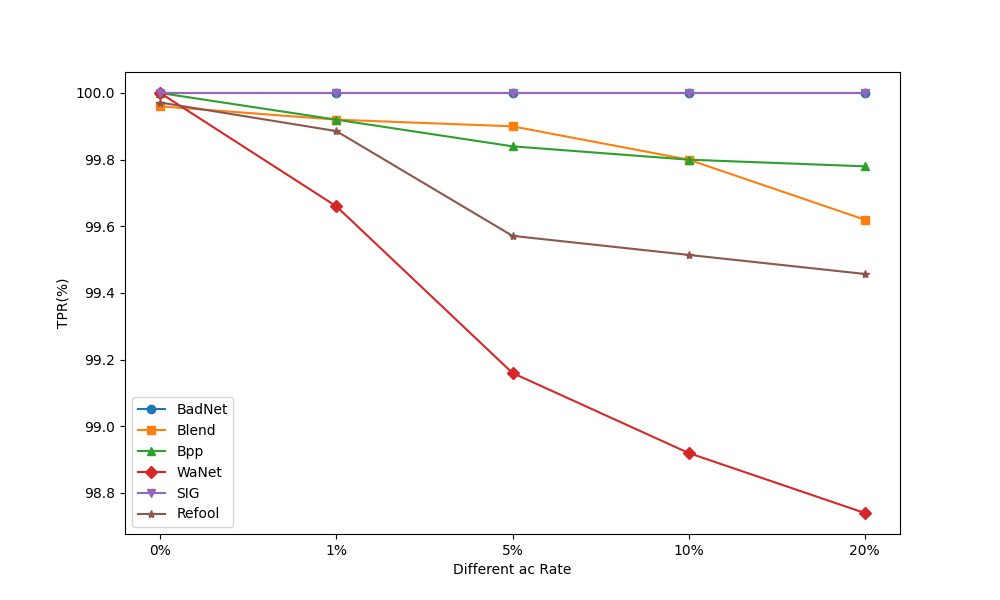}
        \caption{Performance of TPR under fixed $a_p$ in different $a_c$.}
        \label{fig:3}
    \end{minipage}%
    \begin{minipage}[b]{0.5\textwidth}
        \centering
        \includegraphics[width=\linewidth]{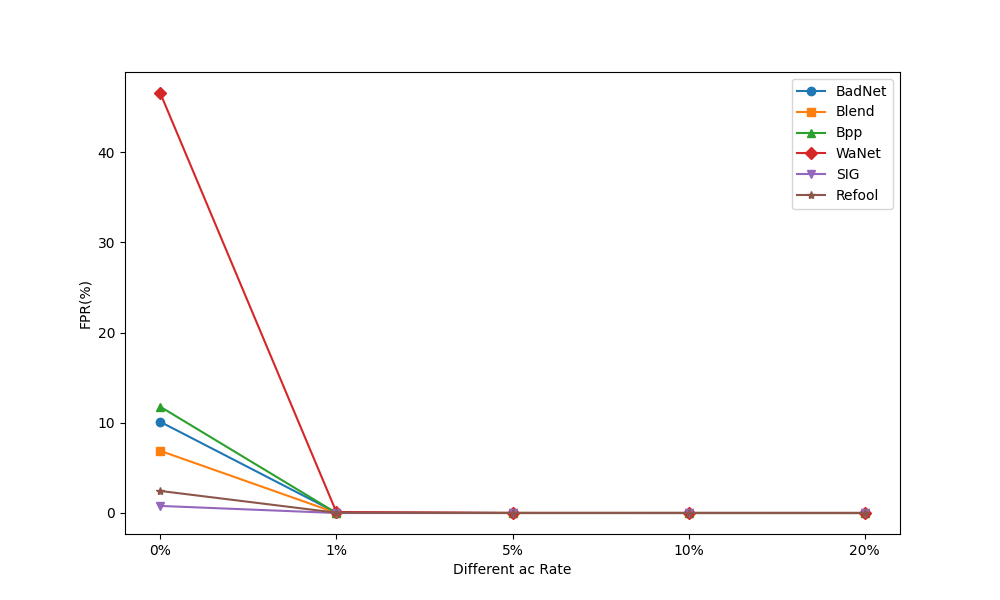}
        \caption{Performance of FPR under fixed $a_p$ in different $a_c$.}
        \label{fig:4}
    \end{minipage}
\end{figure*}

\begin{figure*}[htbp]
    \centering
    \begin{minipage}[b]{0.5\textwidth}
        \centering
        \includegraphics[width=\linewidth]{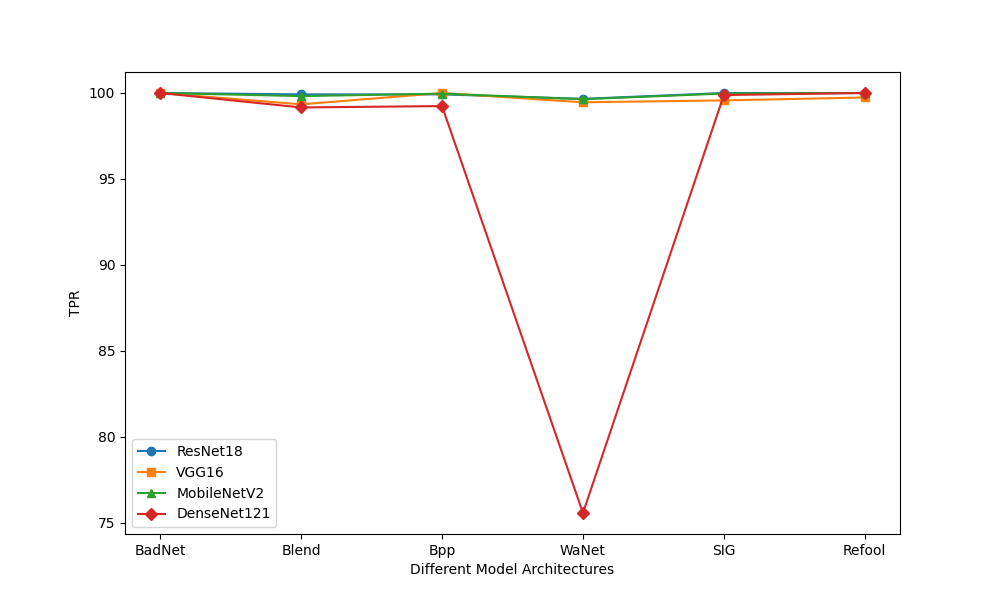}
        \caption{Performance of TPR under different model architectures.}
        \label{fig:5}
    \end{minipage}%
    \begin{minipage}[b]{0.5\textwidth}
        \centering
        \includegraphics[width=\linewidth]{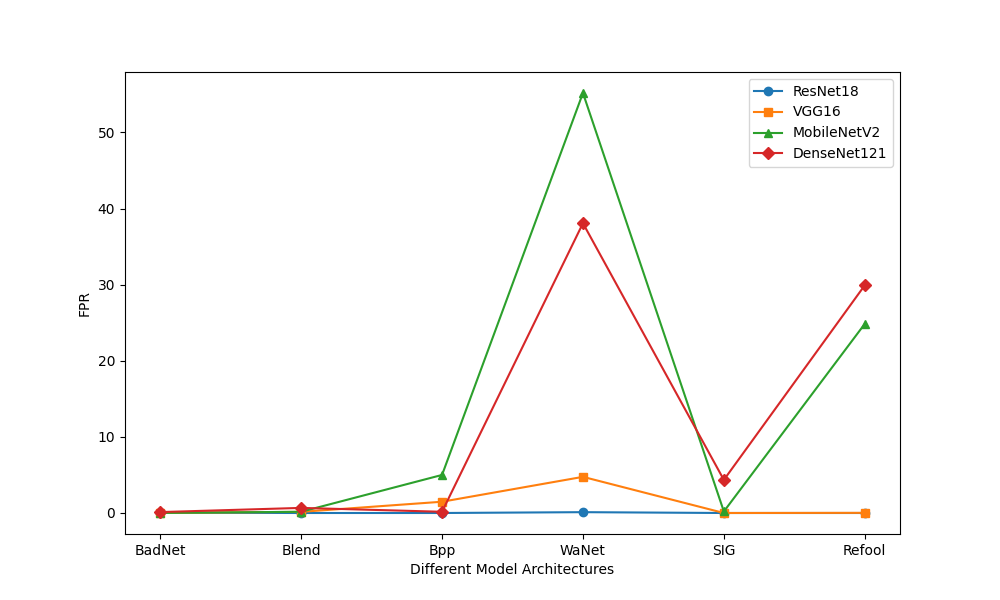}
        \caption{Performance of FPR under different model architectures.}
        \label{fig:6}
    \end{minipage}
\end{figure*}

\section{ABLATION STUDIES}
\label{section5}
\subsection{The effect of isolation rate}
Here, we examine the impact of both $a_c$ and $a_p$ on the efficacy of CBPF's data filtering process. Given that $a_c$ is typically set at 3\% and $a_p$ at 1\% in previous experiments, we vary these parameters to assess the filtering performance.

Initially, we set $a_c$ at 3\% and examine the impact on TPR and FPR while varying the values of $a_p$ (0\%, 1\%, 2\%, 3\%, 5\%, and 10\%). The findings illustrated in Fig.\ref{fig:1} indicate that, with $a_c$ held constant, an increase in $a_p$ results in an elevation of the model's TPR. However, as shown in Fig.\ref{fig:22}, exceeding a $a_p$ value of 3\% leads to a significant increase in FPR. A high FPR implies the exclusion of numerous clean samples alongside the poisoned data, potentially diminishing the model's CAR on datasets with limited data.
This phenomenon is attributed to the presence of partially poisoned data, which is more prevalent when the parameter $a_p$ is larger. This data subset contains a higher proportion of clean data, resulting in the incorporation of benign triggers into the clean dataset after the model associates the new target class with these triggers. Consequently, during the inference phase, some clean images with benign triggers may be misclassified as belonging to the new target class rather than the benign target class.
Therefore, we need to make some trade-offs between TPR and FPR performance, and we find that when $a_p=3\%$ is a better choice, which guarantees a high TPR while also guaranteeing a FPR close to 0\%.

Then, we set $a_p$ to 3\% and systematically varied $a_c$ from 0\% to 20\%. The results are presented in Fig. \ref{fig:3} and Fig. \ref{fig:4}, highlighting the significance of $a_c$ in comparison to other parameters. Specifically, when abstaining from the inclusion of benign triggers and label modifications for isolated clean samples, a high TPR can still be maintained, but at the cost of an increased FPR. 
For instance, WaNet demonstrates a notable FPR of 46.54\%, while badnet and Bpp approach a 20\% FPR. This substantial increase in FPR results in a significant decrease in the model's CAR. In this context, we hypothesize that if benign triggers are not linked to benign labels but are instead incorporated into the poisoned data $a_p$ and the labels are altered to the new target label, it facilitates the classification of samples containing both triggers as the new target class during the inference phase. In this scenario, benign triggers are exclusively found within the poisoned data of $a_p$.
Consequently, these benign triggers essentially serve as a backdoor for the new target class, leading the model to categorize inputs as the new target category once benign triggers are incorporated, regardless of whether the input is poisoned or clean data. Furthermore, a rise in $a_c$ corresponds to a decrease in TPR. Consequently, taking into account the balance between $a_c$'s influence on TPR and FPR, a value of $a_c$ = 1\% is the optimal choice.

\subsection{Different Model Architectures}
In this section, we analyze the filtering effect of the CBPF method on various model architectures, as illustrated in Fig. \ref{fig:5} and Fig. \ref{fig:6}. Our findings indicate that the filtering effect of the CBPF method remains consistent across ResNet18 and VGG16 models, achieving a high TPR while maintaining a FPR below 5\%. However, MobileNetV2 demonstrates a higher TPR but a significantly elevated FPR of 55.18\% when filtering WaNet attacks. This trade-off between filtering effectiveness and preservation of clean samples may potentially compromise the overall accuracy of the model.
Densenet121, as cited in \cite{30denseNet121}, exhibits a TPR of 75.58\% in the context of segregating WaNet's tainted data. This outcome suggests a susceptibility to the inclusion of a high number of contaminated samples within the ostensibly clean dataset utilized for retraining, thereby compromising the efficacy of the defense mechanism. This phenomenon may be attributed to the excessive depth of Densenet121's model, rendering it challenging to train effectively. Moreover, the inherent complexity of WaNet as a backdoor attack further exacerbates the difficulty in learning and subsequently defending against it, thereby contributing to the observed failure of Densenet121's defense strategy.

\begin{table}[htbp]
\centering
\caption{The poisoning filtering effectiveness for six different backdoor attacks at various poisoning rates on CIFAR10.}
\label{different poisoning rates}
\begin{tabular}{l|cclclc}
\hline
\multicolumn{1}{c|}{(\%)} & Attack                  &                      &  & \multicolumn{3}{c}{Poisioning Rate}  \\ \hline
\multirow{2}{*}{}         & \multicolumn{1}{l}{}    & \multicolumn{1}{l}{} &  & 5\%   &  & 20\%                      \\ \cline{2-7} 
                          & \multirow{2}{*}{BadNet}                  & TPR                  &  & 100   &  & 100                       \\
                          &                         & FPR                  &  & 0.02  &  & 0.12                      \\ \cline{2-7} 
                          & \multirow{2}{*}{Blend}  & TPR                  &  & 99.96 &  & 99.97                     \\
                          &                         & FPR                  &  & 0.01  &  & 0.01                      \\ \cline{2-7} 
                          & \multirow{2}{*}{Bpp}    & TPR                  &  & 100   &  & 99.99                     \\
                          &                         & FPR                  &  & 0.2   &  & 0.01                      \\ \cline{2-7} 
CIFAR10                   & \multirow{2}{*}{WaNet}  & TPR                  &  & 99    &  & \multicolumn{1}{l}{99.21} \\
                          &                         & FPR                  &  & 1.95  &  & 0.1                       \\ \cline{2-7} 
                          & \multirow{2}{*}{SIG}    & TPR                  &  & 100   &  & \multirow{2}{*}{/}        \\
                          &                         & FPR                  &  & 0     &  &                           \\ \cline{2-7} 
                          & \multirow{2}{*}{Refool} & TPR                  &  & 99.68 &  & \multirow{2}{*}{/}        \\
                          &                         & FPR                  &  & 3.52  &  &                           \\ \hline
\end{tabular}
\end{table}

\subsection{Different Poisoning Rates}
In this section, we examine the defensive capabilities of the CBPF method across varying levels of poisoning rates. The results presented in Table \ref{different poisoning rates} demonstrate the effectiveness of the method in separating data under poisoning rates of 5\% and 20\%. Furthermore, given that the number of poisonings in a clean label attack can not exceed the number of target classes, we specifically evaluate the impact of the clean label attack at a 5\% poisoning rate. Our findings indicate that the CBPF approach consistently maintains a TPR of over 99\% in accurately filtering poisoned data across different poisoning rates, with the FPR remaining below 1\% for all attacks except Refool. The elevated false positive rate observed in the defense mechanism against Refool attacks at a 5\% poisoning rate may be attributed to the limited presence of uncontaminated data within the subset of images affected by a 3\% poisoning rate. This subset of clean data is subsequently reclassified with the new target label following the introduction of benign triggers, resulting in a correlation between the benign and target labels post-training. Consequently, a minority of clean images containing benign triggers may be erroneously classified by the model during the inference stage.

\subsection{Potential Limitations}
In prior experiments, limitations are identified regarding the uncertainty surrounding the poisoning rate. It is commonly assumed in these experiments that the poisoning rate exceeds a certain threshold, denoted as $a_p$. However, this assumption may not accurately reflect real-world scenarios. When the poisoning rate is lower than $a_p$, a minimal amount of data is poisoned during the initial filtering stage, including some clean samples. Consequently, this results in a higher loss of clean samples in the process of Composite Backdoor Poison Filtering (CBPF). In this context, how to obtain high poisoning-rate data without knowing the poisoning rate will serve as our future work.

\section{Conclusions}
\label{section7}
In this paper, we introduce a novel defense framework called CBPF that utilizes backdoor methods to filter out poisoning samples. Our analysis reveals notable disparities in the outputs of poisoned and clean samples when processed by the contaminated model. Leveraging this finding, we segregate a limited subset of tainted and untainted samples. We employ benign triggers on the isolated clean sample subset and modify their benign labels accordingly. Likewise, for the poisoned data, we introduce benign triggers and reassign them to new target labels.
In the process of inference, benign triggers are applied to all input data, resulting in the model predicting clean samples with benign labels in the presence of benign triggers, and poisoned samples with altered target labels when benign triggers are present. Extensive experimentation has shown that CBPF is capable of effectively defending against attacks with minimal impact on performance when processing clean data. Furthermore, the high accuracy of our method in identifying and filtering out poisoned data ensures that the remaining clean data is largely free from contamination, enhancing the portability of the CBPF. This allows for flexible integration of our approach with other backdoor defense methods to achieve superior defense effectiveness.

\section*{Acknowledgements} This work is supported by the National Natural Science Foundation of China (Grant No.61602408), and Zhejiang Provincial Natural Science Foundation of China under Grant (Nos.LY19F020005).


\ifCLASSOPTIONcaptionsoff
  \newpage
\fi



\bibliographystyle{IEEEtran}
\bibliography{de}{}
\end{document}